\documentclass[prb,twocolumn,superscriptaddress,floatfix]{revtex4-1}
\usepackage{amsfonts,amsmath,graphicx,amssymb,bm}
\usepackage[pdftex,plainpages=false,colorlinks=true,linkcolor=blue, citecolor=blue, urlcolor=blue]{hyperref}
\pdfoutput=1

\usepackage[x11names,svgnames]{xcolor}

\newcommand\Bv{\mathbf{B}}
\newcommand\Bh{\mathbf{\hat B}}
\newcommand\Gv{\mathbf{G}}
\newcommand\Mv{\mathbf{M}}
\newcommand\Tv{\mathbf{T}}
\newcommand\kv{\mathbf{k}}
\newcommand\Kv{\mathbf{K}}
\newcommand\rv{\mathbf{r}}
\newcommand\kvt{\mathbf{\tilde k}}
\newcommand\bv{\mathbf{b}}
\newcommand\ev{\mathbf{e}}
\newcommand\sigmav{\bm{\sigma}}
\newcommand\Sigmav{\Sigma}
\newcommand\er{\mathrm{e}}
\newcommand\dr{\mathrm{d}}
\newcommand\PB{R}

\begin{document}

\title{Quarter-filled Kitaev-Hubbard Model: A Quantum Hall State in an Optical Lattice}

\author{S. R. Hassan, Sandeep Goyal, R. Shankar}
\affiliation{The Institute of Mathematical Sciences, C.I.T. Campus, Chennai 600
113, India}
\author{David S\'en\'echal}
\affiliation{D\'epartment de Physique and RQMP, Universit\'e de Sherbrooke,
Sherbrooke, 
Qu\'ebec, Canada J1K 2R1}

\date{\today}

\begin{abstract} 

We analyze the physics of cold atoms in honeycomb optical lattices with on-site
repulsion and spin-dependent hopping that breaks
time reversal symmetry. Such systems, at half filling and large on-site
repulsion, have been proposed as a
possible realization of the Kitaev model. The spin-dependent hopping breaks
the spin degeneracy and, if strong-enough, leads to four
non-overlapping bands in the non-interacting limit. These bands carry
nonzero Chern number and therefore the non-interacting system has nonzero
angular momentum and chiral edge states at 1/4 and 3/4 filling. We have
investigated the effect of interactions on a quarter-filled system using the 
variational cluster perturbation theory and found that the critical spin-dependent hopping
that separates the metal from the quantum Hall state is affected
by interactions.

\end{abstract}


\maketitle

\section{Introduction}

Condensed matter systems with topological quantum order have been studied for
several decades now. Initially, the focus was on quantum Hall states,
but it was later realized that topological effects arise from
fermionic band structures too.\cite{Haldane:1988fk} The effect of
interactions in such topological systems is of current interest. 

The Kitaev model on the honeycomb lattice is a quantum spin-$\frac12$ model
that has such topological order.\cite{Kitaev:2006fk} There have been recent
proposals \cite{Duan:2003dq, Zhang:2007cr, Dusuel:2008fk, *Vidal:2008uq} to
realize this model in an optical lattice, using two-state bosonic or fermionic
atoms with spin-dependent hopping on a honeycomb lattice with on-site
repulsion.

\begin{figure}[h]
\includegraphics[scale=0.9]{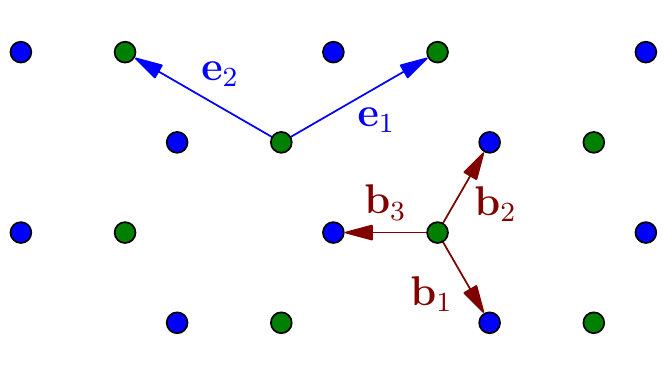}
\caption{(Color online) The honeycomb lattice, the NN vectors $\bv_a$ and 
the lattice basis $\ev_{1,2}$. The A and B sublattices are represented by green and blue dots, respectively.}
\label{fig:vectors}
\end{figure}

The system proposed by Duan et al.\cite{Duan:2003dq} consists of atoms trapped
in a two dimensional honeycomb optical lattice. They have suggested a scheme
by which the potential barriers between the neighboring minima of the trapping
potential can be made spin and direction dependent. The scheme is to choose two
atomic states with slightly different energies to be the two effective spin
states. These two states are coupled to a common excited level by shining two
phase locked laser beams with frequencies that are slightly blue detuned to the
corresponding transitions. The laser beams can be chosen such that they do not
change positions of the minima of the trap potential. They will however induce
a spin dependence in the potential barrier between two neighboring minima in
the direction of the laser beams. The form and strength of the spin dependent
potential would depend on the relative phase and amplitudes of the two beams.
Thus by applying three pairs of laser beams in the three tunneling directions
with different relative phases and amplitudes, a spin and direction dependent
potential barrier can be engineered. Such a trapping potential barrier will
result in spin and direction dependent hopping matrix elements in an effective
tight-binding model that describes the low energy dynamics of the system.

We therefore study an effective tight-binding model on a honeycomb lattice.
The unit cell positions span a 
triangular lattice and are given by $i_1\hat\ev_1+i_2\hat\ev_2$, where the 
basis vectors are taken to be $\hat\ev_{1(2)}=(\pm\frac32,\frac{\sqrt3}2)$ 
(see Fig.~\ref{fig:vectors}).
We label sites by $i = (i_1,i_2,s)$ where $s=A,B$ labels the sublattice. 
We denote the three directions linking a site on the A sublattice with its three neighbors on the B sublattice by $\bv_a$, $a=1,2,3$ (Fig.~\ref{fig:vectors}).
The nearest-neighbor pairs along the $\bv_a$ direction are denoted by 
$\langle ij\rangle_a$. 
The Hamiltonian is then
\begin{equation}\label{eq:Kitaev-Hubbard}
H=\sum_{\langle ij\rangle_a}
\left\{\frac12 c^\dagger_{i\alpha}(t\delta_{\alpha\beta}+t'\sigma^a_{\alpha\beta}) c_{j\beta}+\mathrm{H.c.}\right\}
+U\sum_i~n_{i\uparrow}n_{i\downarrow}
\end{equation}
where $c^\dagger_{i\alpha}$ creates a fermion of spin projection $\alpha$ at site $i$, and $n_{i\alpha} = c^\dagger_{i\alpha}c_{i\alpha}$ is the number operator.
$\sigma^a$ ($a=1,2,3$) are the Pauli spin matrices:
\begin{equation}
\sigma^1=\left(\begin{array}{cc}0&1\\1&0\end{array}\right),
~~\sigma^2=\left(\begin{array}{cc}0&-i\\i&0\end{array}\right),
~~\sigma^3=\left(\begin{array}{cc}1&0\\0&-1\end{array}\right)
\end{equation}

Model \eqref{eq:Kitaev-Hubbard}, which we call the Kitaev-Hubbard model, reduces to the Kitaev spin model at half-filling and large on-site repulsion $U$.\cite{Duan:2003dq, Zhang:2007cr,Dusuel:2008fk,*Vidal:2008uq,Jordens:2008nx}
We have recently shown\cite{Hassan:2013fk} that at half filling, this model
supports a stable algebraic spin liquid at intermediate values of
the on-site repulsion. 
In this paper, we will rather study the same model at quarter filling.

At $t'=0$, the model reduces to the simple spin-invariant, nearest-neighbor (NN)
Hubbard model, as relevant to graphene, and is time-reversal invariant.
At $t'=t$, the one-body part of the Hamiltonian is a combination of the
projection operators $\frac12(1+\sigma^a)$. 
Thus, only particles spin polarized in the 
$a$ direction can hop along the $a$ bonds. 
The term proportional to $t'$ is qualitatively
different from the usual spin-orbit term as it breaks time reversal
symmetry. 

To see this, let us denote the hopping matrix between the nearest 
neighbor sites in direction $a$ by $T_a= t+t'\sigma^a$. 
Time reversal invariance requires $T_a= \sigma^2 T_a^*\sigma^2$. 
However, we have $\sigma^2 T_a^*\sigma^2=t-t'\sigma^a\ne T_a$.

The physical origin of this lies in the asymmetry of the amplitudes and phases
of the two laser beams coupling the two spin components. If the phases and
amplitudes of these two beams are interchanged, we find $t'\rightarrow
-t'$, i.e., the time-reversed model. It is interesting to note that a
time-reversal invariant hopping term of the form $\frac12
c^\dagger_i\left(t+it'\sigma^a\right) c_j$, which is of the form of a standard
spin-orbit coupling, does not lead to the Kitaev model at half filling in the
large-$U$ limit.

Topological effects in electronic bands were highlighted in the seminal paper 
of Thouless {\em et al.\/} \cite{Thouless:1982fk} where the quantized Hall
conductivity 
was expressed as the sum of the Chern numbers of the occupied bands (when the
Fermi level lies in a gap).
The Chern number and hence the quantized Hall conductivity was later identified
with
the number of chiral edge channels at the Fermi level by
Hatsugai.\cite{Hatsugai:1993ve} A nonzero Chern number can only occur in a
system which is not time reversal invariant. However, this does not necessitate
an external magnetic field: Haldane constructed a tight-binding model on a
honeycomb lattice with next-NN hopping terms that are not
time-reversal invariant.\cite{Haldane:1988fk} In that model the two bands carry
Chern
numbers equal to $\pm1$ and an anomalous Hall effect occurs when
the bands are partially filled.\cite{Haldane:2004qf} The Hall conductivity
is quantized when the Fermi level lies in the gap. More recently, the
topology of time reversal invariant electronic bands with spin-orbit
couplings has been extensively studied in the physical context
of the spin Hall effect, leading to the discovery of topological
insulators.\cite{Hasan:2010fk,Qi:2011uq} 
Effects of interactions on such systems with spin-orbit coupling have also been studied.\cite{Hohenadler:2011fk}
The effect of time-reversal-breaking spin-dependent hopping, to our knowledge, has not been studied so far. 

In the rest of the paper, we concentrate on Model (\ref{eq:Kitaev-Hubbard}) at
$\frac14$ (or $\frac34$) filling and show that it is in an anomalous quantized
Hall state in a region of the ($t'$-$U$) plane.
At $U=0$, the four
bands carry Chern numbers equal to $\pm 1$. At large enough $t'$, they do
not overlap, leading to quantized anomalous Hall states at fillings
$\frac14$ and $\frac34$. We analyze the corresponding chiral edge state
structure and propose a way to detect them. We then study the model at
$U>0$ using Cluster Perturbation Theory \cite{Senechal:2000} (CPT) and
the Variational Cluster Approximation.\cite{Potthoff:2003b}
We show that, for an interval of $t'$, it is possible to go from the
metallic state to the Hall state upon increasing $U$, a transition that
could be observed in ultra-cold atom systems.

Previous work on realizing quantum Hall states in cold atom systems devised schemes to simulate a magnetic field acting on the neutral atoms.
For instance, Baranov et al.\cite{Baranov:2005uq} exploit the analogy between a rotating frame and a magnetic field. 
On the other hand, Sorensen et al.\cite{Sorensen:2005kx} use an oscillating quadrupole field to induce phases in the hopping elements of the effective tight-binding model, which is exactly the effect of a magnetic field on a charged particle hopping on the lattice.
By contrast, the physics of the Kitaev-Hubbard model (\ref{eq:Kitaev-Hubbard}) is similar to that of the Haldane model, where the nonzero Chern number comes from band effects. Whereas the Haldane model has complex next-nearest-neighbor hopping on the honeycomb lattice, which breaks time-reversal symmetry
and gives rise to nonzero Chern number, our model has only nearest-neighbor
hopping; the nonzero Chern number is rather due to the nature of the time-reversal-breaking, spin-dependent hopping.

\section{The non-interacting limit}

In this section we solve model (\ref{eq:Kitaev-Hubbard}) in the $U=0$ limit. Hereinafter, we set 
$t=1$. Thus all energies are in units of $t$.
We define the Fourier transform
\begin{equation}
c_{\kv s\alpha}=\sum_{i_1,i_2} \er^{i\kv\cdot(i_1\hat\ev_1+i_2\hat\ev_2)}c_{i_1,i_2,s,\alpha}
\end{equation}
Let $P_a=\frac12(1+t'\sigma_a)$ and $k_{1(2)}\equiv\kv\cdot\hat\ev_{1(2)}$.
The Hamiltonian in momentum space can then be written as
\begin{equation}\label{eq:hammom}
H=\sum_\kv
\left(\begin{array}{cc}c^\dagger_{\kv,1}&c^\dagger_{\kv,2}\end{array}\right)
\left(\begin{array}{cc}0&\Sigmav(\kv)\\
\Sigmav^\dagger(\kv)&0\end{array}\right)
\left(\begin{array}{c} c_{\kv,1}\\c_{\kv, 2}\end{array}\right)
\end{equation}
where $\Sigmav(\kv)=P_3+P_1\er^{ik_2}+P_2\er^{-ik_1}$ is a $2\times2$ matrix in spin space.
Let us write an eigenvector of the above matrix as $(\phi,\psi)$, where $\phi$ and $\psi$ are two-component spinors.
The eigenvalue equation is then
\begin{equation}
\Sigma(\kv)\psi = \epsilon(\kv)\phi \qquad \Sigma^\dagger(\kv)\phi = \epsilon(\kv) \psi
\end{equation}
which, after eliminating $\phi$, reduces to $\Sigma^\dagger(\kv)\Sigma(\kv)\psi = \epsilon^2(\kv)\psi$.
Thus the single-particle spectrum is completely determined in terms of the spectrum of the positive semi-definite matrix
\begin{equation}
\Sigmav^\dagger\Sigmav = f^*f+\frac34 (t')^2
+\frac{t'}{2}\Bv\cdot\sigmav
\end{equation}
where
\begin{align}
f &= \frac12(1+\er^{ik_1}+\er^{-ik_2})\nonumber\\
B_1 &= 1-t'\sin k_1+\cos k_2+\cos k_3 \nonumber\\
B_2 &= 1+\cos k_1-t'\sin k_2+\cos k_3 \nonumber\\
B_3 &= 1+\cos k_1+\cos k_2-t'\sin k_3 
\end{align}
The eigenvalues of $\Bv(\kv)\cdot\sigmav$ are $\pm|\Bv(\kv)|$, and the corresponding normalized eigenvectors are
\begin{equation}\label{eq:spinors}
\psi^- = 
\begin{pmatrix} \er^{i\phi}\sin\theta/2 \\ -\cos\theta/2
\end{pmatrix}\qquad
\psi^+ = 
\begin{pmatrix} \er^{i\phi} \cos\theta/2\\ \sin\theta/2
\end{pmatrix}
\end{equation}
where $\theta$ and $\phi$ are the polar and azimutal angle defining the vector $\Bv = |\Bv|(\sin\theta\,\cos\phi, \sin\theta\,\sin\phi, \cos\theta)$.
$\psi^\pm$ are also the eigenvectors of $\Sigma^\dagger(\kv)\Sigma(\kv)$ with eigenvalues
\begin{equation}
\epsilon^2_{\pm}(\kv)=f^*f+\frac34(t')^2
\pm\frac{t'}{2}\vert\Bv(\kv)\vert
\end{equation}
Since $\Sigma^\dagger(\kv) = \Sigma(-\kv)$, it is a simple matter to see that the second spinors $\phi^\pm(\kv)$ are identical to $\psi^\pm(-\kv)$, up to a phase factor.
The four-component vectors
\begin{equation}\label{eq:phidef}
\Phi^{pp'}(\kv)=
\frac{1}{\sqrt 2}\left(\begin{array}{c}\psi^p(\kv)\\
p e^{i\chi(\kv)}\psi^{p'}(-\kv)
\end{array}\right)
\end{equation}
where $p=\pm$ and $p'=\pm$, 
are then the eigenvectors of the single-particle Hamiltonian defined 
by Eq.~(\ref{eq:hammom}) with eigenvalues $p\epsilon_{p'}(\kv)$. 
The phase factor $\chi(\kv)$ we will discussed below.

\begin{figure}[h]
\includegraphics[width=0.9\hsize]{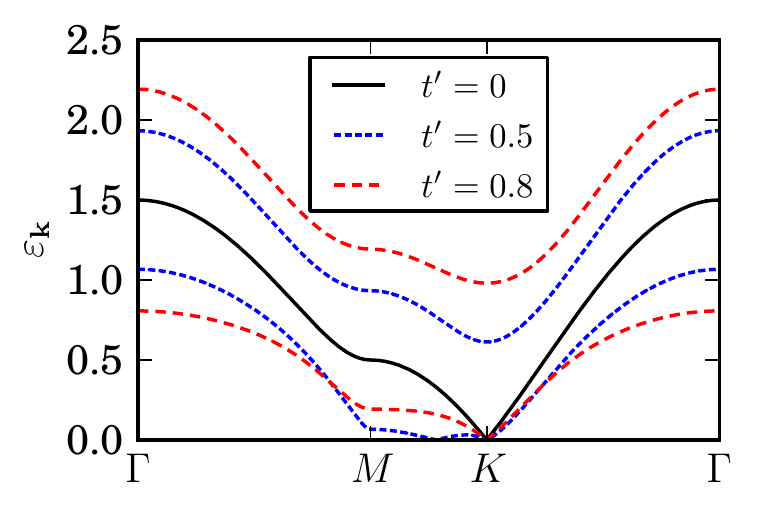}
\caption{(Color online) The dispersion relation $\epsilon_\pm(\kv)$ along a high-symmetry
path in the Brillouin zone. The negative energy bands are symmetrically located underneath 
the horizontal axis and do not touch the upper bands. $K$ is the Dirac point, and 
$M=(K+K')/2$ is the mid-point between the two Dirac points. Note the new Dirac point 
at $t'=0.5$, about midway between $K$ and $M$. At $t'=0$, $\epsilon_+$ and $\epsilon_-$ are degenerate.}
\label{fig:bands12}
\end{figure}

At $t'=0$, we recover the graphene spectrum:
There are two spin-degenerate bands touching each other at two distinct Dirac
points ($\Kv=(2\pi/3,2\pi/3\sqrt3)$ and $\Kv'=(-2\pi/3,-2\pi/3\sqrt3)$) at the zone boundary.
At any nonzero value of $t'$, the spin degeneracy is broken.
Two of the bands, $\epsilon_-$ and $-\epsilon_-$, continue to touch each
other at two Dirac points whereas the other two develop a gap.
The $\epsilon_+$ and $\epsilon_-$ bands
overlap until a critical value $t'_c$ of $t'$ given by $\sqrt 6-\sqrt 3=0.717$.
For $t'> t'_c$, a finite gap appears.
This is illustrated in Fig.~\ref{fig:bands12}.

At $t'=\sqrt{3}$, the middle two bands, $\epsilon_{-}$ and $-\epsilon_{-}$,
touch each other at $k=0$.
For $t'>\sqrt{3}$, the point of contact splits into six more Dirac points, for a total of eight Dirac points.

\section{Chern numbers and Orbital Magnetization}
\label{cnom}

The Pancharatnam-Berry (PB) curvature for each band is given by
\begin{equation}
\PB^{pp'}(\kv)=\frac{\epsilon^{\mu\nu}}{8\pi i}
\left(\partial_\mu\Phi^{pp'}(\kv)^\dagger\partial_\nu\Phi^{pp'}(\kv)-\mathrm{H.c.} \right)
\end{equation}
where $\mu,\nu=1,2$ are two orthogonal directions in the Brillioun zone.
$\epsilon^{\mu\nu}$ is the two-dimensional Levi-Civita tensor.
Hereinafter, we use the Einstein convention of summing over repeated indices.
From Eqs~\eqref{eq:phidef} and \eqref{eq:spinors} and the discussion in Ref.~\onlinecite{Xiao:2010tg}, it follows that
\begin{eqnarray}\label{eq:PB}
\nonumber
\PB^{pp'}(\kv)&=&p'\frac{1}{2}\left(b(\kv)+b(-\kv)\right)
+\frac{1}{2}\epsilon^{\mu\nu}\partial_\mu\partial_\nu\chi(\kv)\\
b(\kv)&=&\frac{\epsilon^{\mu\nu}}{8\pi}
\Bh(\kv)\cdot\partial_\mu\Bh(\kv)\times\partial_\nu\Bh(\kv)
\end{eqnarray}
It is not difficult to prove that $|\Bv|\ne 0$ for all points in the 
Brillouin zone, {\em including the Dirac points}, for any nonzero
value of $t'$.
Thus $\PB^{pp'}(\kv)$ is well defined throughout the Brillouin zone.
However the phase factor $\chi(\kv)$ is multi-valued at the Dirac points
and gives rise to the delta function contribution to the PB curvature
there with strength $\pm\pi$. Since we are concentrating on the quarter
(three quarter) filled case in this paper and there are no Dirac points
in the lowest and highest energy bands, we can afford to ignore this 
phase $\chi(\kv)$.

The integral $\int_\kv b(\kv)$ is the number of times the vector $\Bh(\kv)$ sweeps the unit sphere as $\kv$ covers the whole Brillouin zone.
We have numerically verified that it is equal to 1.
Therefore, the highest and lowest energy bands with energies, $\pm\epsilon_{+}$, 
have Chern numbers +1 whereas the middle two bands with energies,
$\pm\epsilon_{-}$, have Chern numbers $-1$.

When $t'>0.717$ and the bands are non-overlapping, the half-filled state has
total Chern number zero but the quarter and three-quarter filled states have
Chern numbers equal to $\pm 1$ respectively.
These states are thus analogous to quantum Hall states with quantized Hall
conductivity $\sigma_H=\pm 1$.
At $t'<0.717$, the bands overlap and at quarter filling the Fermi level passes
through the $-\epsilon_-$ bands.
The system is then analogous to an anomalous Hall conductor with the ``Hall
conductivity" equal to the PB curvature integrated over the
occupied states.\cite{Haldane:2004qf}

\begin{figure}[h]
\includegraphics[width=0.9\hsize]{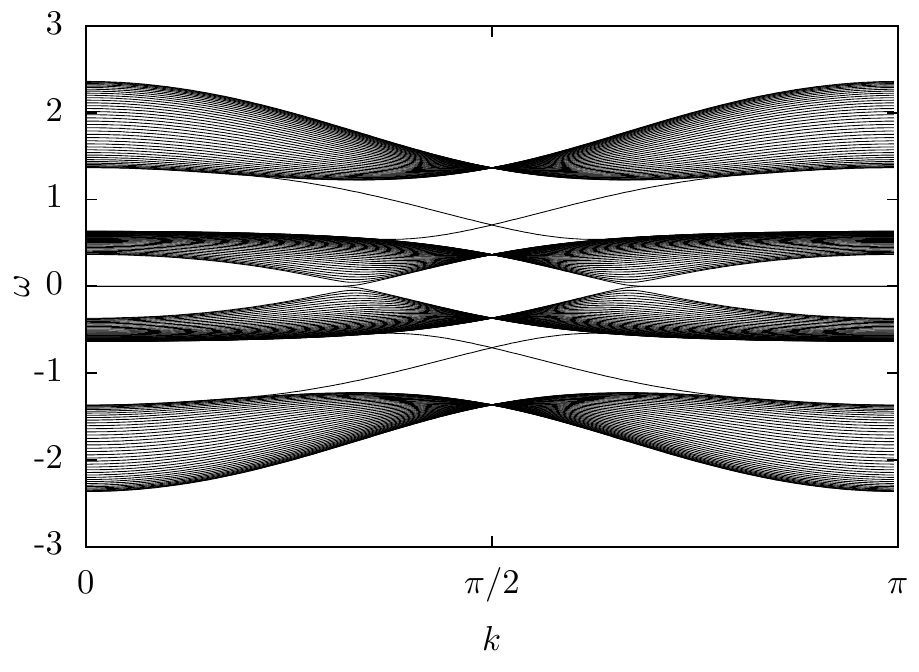}
\caption{The spectrum for an open tube of circumference $L=60$ with zig-zag
edges, for $t'=t$.
Note the edge states that cross the gaps between the bands.}
\label{fig:edges}
\end{figure}

In the regime $t'>0.717$, we expect chiral edge states in the gap between 
the $\epsilon_+$ and $\epsilon_-$ bands likewise between the 
$-\epsilon_-$ and $-\epsilon_+$ bands.
Exact diagonalizations in finite cylindrical
geometry and zig-zag edges (Fig.~\ref{fig:cluster}B) confirm this.
Figure \ref{fig:edges} shows the corresponding spectrum, with non-dispersive, zero energy edge states, as in graphene.
There are also chiral edge states between the top two and the bottom two bands.

In a system of neutral atoms, the Hall conductivity is not easily 
measurable. However, the non-trivial topology also manifests itself in the
orbital angular
momentum, which is easier to measure then.
The orbital magnetization of Bloch particles of the band $pp'$ is given 
by \cite{Shi:2007kl,Xiao:2010tg,Xiao:2005hc,Thonhauser:2005ij,Xiao:2006bs}
\begin{eqnarray}\label{eq:angularm}
\Mv_{{pp'}}& = &\frac{e}{2\hbar}\int_{p\epsilon_{p'}(\kv)\le\mu}
\frac{\dr^2k}{(2\pi)^2}\nonumber \\
&\times& \langle{\partial}_\kv\Phi^{pp'}|(H_\kv +p\epsilon_{p'}(\kv)-2\mu) 
|{{\partial}_\kv}\Phi^{pp'}\rangle\mbox{,}
\end{eqnarray}
where $\mu$ is the chemical potential (Fermi energy) and 
$\vert\Phi_{pp'}\rangle$ are the single particle eigenvectors.
In the metallic case, Eq.~(\ref{eq:angularm}) provides a $\mu$-dependent
magnetization, as it should.
In the insulating case, when $\mu$ is varied in the gap, $\Mv$ changes linearly
only if the Chern invariant is nonzero, and remains constant otherwise. 
Eq.~(\ref{eq:angularm}) is related -- though not identical -- to the anomalous
Hall conductivity.

\begin{figure}[h]
\includegraphics[width=0.8\hsize]{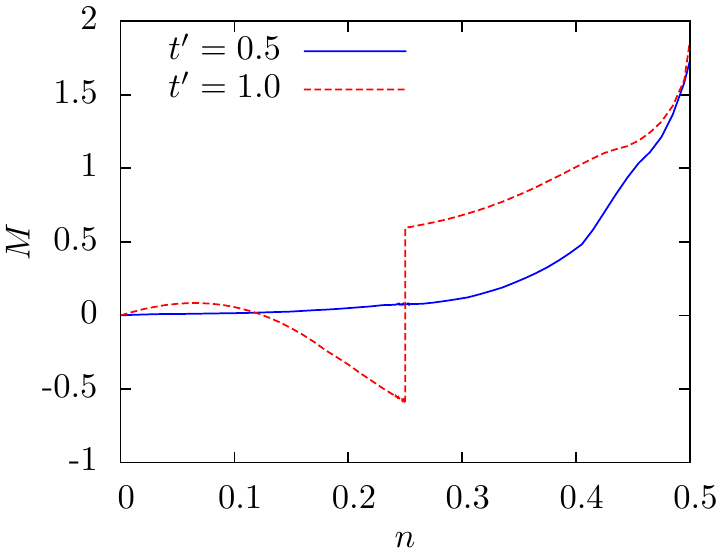}
\caption{(Color online) Orbital magnetization of the model as a function of the
filling $n$
 for $t'$=0.5,1.0}
\label{fig:ahe}
\end{figure}

We computed the orbital magnetization as a function of the filling $n$ by
Eq.~(\ref{eq:angularm}) in the metallic region at $t' =0.5$ and in the
insulating region at $t'=1.0$. 
The behavior of the orbital magnetization as the Fermi energy varies from the
bottom of the lowest band to the top of the second band is displayed in
Fig.~\ref{fig:ahe}. 
The orbital magnetization in the insulating phase at $t' =1$ shows a 
discontinuity at $n=0.25$, because the integral of the PB 
curvature over the Brillouin zone is nonzero and quantized.
The anomalous quantized Hall conductivity is the ratio of the discontinuity 
in the orbital magnetization to $E_g/(2e)$, where $E_g$ is the gap 
between the first and the second band. 

Thus, the nonzero Hall conductivity manifests itself as a nonzero orbital angular
momentum. In other words, there is a rotating condensate in a static
optical trap. The above calculation corresponds to a system with a sharp edge.
However, the same physics holds for a slowly varying confinement potential as
in a realistic optical trap. To see this, consider the
semiclassical equations derived by Sundaram and Niu \cite{Sundaram:1999fk} for a 
wave packet of an atom restricted to the lowest band in the presence of a
slowly varying confining potential. For wave packets whose width in real space
is much larger than the lattice spacing, the dynamics of the mean wave vector
${\bf k}$ and the mean position $\rv $ are governed by the equations
\cite{Sundaram:1999fk}
\begin{eqnarray}
\dot x^\mu&=&-\frac{\partial\epsilon_-({\bf k})}{\partial k_\mu}
+\epsilon^{\mu\nu}\PB^{--}({\bf k})\dot k_\nu\\
\dot k_\mu&=&-\frac{\partial V(\rv )}{\partial x^\mu}
\end{eqnarray}
where $V(\rv )$ is the confining potential, $-\epsilon_-({\bf k})$
and $\PB^{--}({\bf k})$ are the kinetic energy and the 
PB field of the lowest band.
For a circularly symmetric confining potential, the force on the atoms
will be radial and we put $-{\bf \nabla} V=F(r){\bf \hat r}$, where 
$\rv =r{\bf \hat r}$.
Thus the velocity of the atom, ${\bf v}=\dot\rv $ is,
\begin{equation}
v^\mu=-\frac{\partial\epsilon_-(k)}{\partial k_\mu}
+\epsilon^{\mu\nu}\frac{x_\nu}{r} F(r)\PB^{--}(k)
\end{equation}
Within the semiclassical approximation, the velocity field of the many particle
system is obtained from the above equation by summing over all the occupied
wave vectors. The second term, the so called anomalous velocity, is tangential 
and its magnitude is proportional to the total PB curvature
integrated over the occupied states. As we have shown earlier, this is 
nonzero in the lowest band. Thus a slowly varying static confining potential 
will also induce the fermionic condensate to rotate.

\section{Interactions}

Thus far, the calculations we have presented were obtained in the 
non-interacting limit ($U=0$). We now study the model as
a function of the Hubbard interaction $U$ to determine the region
in the parameter space ($t'-U$) where the topological effects persist.

In the presence of interactions, the Chern number may still be calculated in principle, using the following expression:\cite{Volovik:2001vn}
\begin{equation}\label{eq:Chern}
N = \frac1{24\pi^2} \int\dr\omega\dr^2k\,\epsilon^{\mu\nu\lambda}\, \mathrm{tr}\left\{ G\partial_\mu G^{-1} G\partial_\nu G^{-1} G\partial_\lambda G^{-1}\right\}
\end{equation}
where $G$ stands for the Green function, the integral is taken over frequency 
and momentum and the trace is taken over spin and band indices. Greek indices are space-time (wave vector-frequency) indices running from 0 to 2 with the 
convention that $k_0=\omega$. Also, $\epsilon^{\mu\nu\lambda}$ is the three 
dimensional Levi-Civita tensor. 
In the non-interacting case, this expression coincides with the integral of the PB curvature over occupied states of the Brillouin zone.
In the interacting case, the Green function will be deformed, but as long as a gap persists at the Fermi level, the Chern number should not be affected, as it is robust against smooth deformations that do not introduce or remove any poles in $G$.
The Chern invariant will also
manifest itself as chiral edge states in the interacting system.
We therefore address two questions: What is the region of parameter space
where the gap persists? Are there chiral edge states in that region?

We study the interacting theory using Cluster Perturbation 
Theory \cite{Senechal:2000} (CPT) and the Variational Cluster 
Approximation.\cite{Potthoff:2003b}

CPT is an approximation scheme for the one-electron Green function $\Gv(\omega)$ within Hubbard-like models.\cite{Senechal:2000, Senechal:2002,Senechal:2011vn}
It proceeds by dividing the infinite lattice $\gamma$ into a superlattice $\Gamma$ of identical clusters of $L$ sites each.
The lattice Hamiltonian $H$ is written as $H = H_c + H_T$, where $H_c$ is the cluster Hamiltonian, obtained by severing the hopping terms between different clusters, which are put into $H_T$.
Let $\Tv$ be the matrix of inter-cluster hopping terms and ${\Gv^{c}}(\omega)$ the exact Green function of the cluster.
Because of the periodicity of the superlattice, $\Tv$ can be expressed as a function of the reduced wave vector $\kvt$ and as a matrix in site indices within the cluster: $T_{mn}(\kvt)$ ($m$ and $m$ will be used to denote collectively lattice site and spin). Likewise, ${\Gv^{c}}$ is a matrix in cluster site indices only, since all clusters are identical: $G^c_{mn}(\omega)$. Thus, hopping matrices and Green functions in what follows will be $\kvt$-dependent matrices of order $L$, the number of sites within each cluster.
The CPT approximation for the Green function is
\begin{equation}\label{eq:CPT1}
\Gv^{-1}(\kvt,\omega)={\Gv^{c}}^{-1}(\omega)-\Tv(\kvt)
\end{equation}
In practice ${\Gv^{c}}(\omega)$ is calculated numerically by the Lanczos method and the cluster must be small enough for this to be possible.
Because the lattice tiling breaks the original translation invariance of the lattice, a prescription is needed to restore the translation invariance of the resulting Green function.
The CPT prescription for this periodization is
\begin{equation}
G(\kv,\omega) = \frac1{L}\sum_{m,n} \er^{-i\kv\cdot(\rv_m-\rv_n)} G_{mn}(\kv,\omega)
\end{equation}
where now $\kv$ belongs to the Brillouin zone of the original lattice.
This formula is exact in both the strong $(t\to 0)$ and the weak ($U\to 0$) coupling limits.

Once the approximate interacting Green function can be calculated, various quantities can be calculated, such as the electron density $n(\mu)$ as a function of chemical potential, or the spectral function $A(\kv,\omega)=-\frac{1}{\pi}\mathrm{Im}\, G(\kv,\omega)$ and its integral over wave vectors, the density of states $N(\omega)$.

The Variational Cluster Approximation (VCA) is an extension of CPT in which parameters of the cluster Hamiltonian $H_c$ may be treated variationally, according to Potthoff's Self-Energy Functional Theory (SFT).\cite{Potthoff:2003b,Potthoff:2012fk}
In particular, it allows the emergence of spontaneously broken symmetries and provides an approximate value for the system's grand potential $\Omega$. Technically, VCA proceeds by minimizing the following quantity:
\begin{equation}
\Omega(h) = \Omega_c(h) - \!\int\frac{\dr\omega}{\pi}\frac{\dr^2k}{(2\pi)^2}\sum_\kvt \ln\det\left[\mathbf{1}\!-\!\Tv(\kvt)\Gv(\kvt,i\omega)\right]
\end{equation}
where $\Omega_c(h)$ is the grand potential of the cluster alone (obtained by exact diagonalization).
The integral over frequencies $k_0$ is carried over the imaginary axis, and $h$ denotes collectively the parameters of the cluster Hamiltonian $H_c$ that are treated variationally; these must be the coefficients of one-body operators.
At the optimal value $h^*$, $\Omega(h^*)$ is the best estimate of the system's grand potential; in particular, its derivative
$\partial\Omega/\partial\mu=-n$ gives us a reliable estimate of the electron density.
VCA provides estimates of order parameters, much like mean-field theory, but is quite superior to it because the Hamiltonian remains fully interacting (no factorization of the interaction) and spatial correlations are treated exactly within the cluster. It has been widely applied to systems of strongly correlated electrons, mostly to investigate ordered phases.
For a general review, see Ref.~\onlinecite{Potthoff:2012fk}, and Refs~\onlinecite{Dahnken:2004,Senechal:2005,Sahebsara:2006bs,Sahebsara:2008fv,Arrigoni:2009fe} for examples of its use.

\begin{figure}[h]
\includegraphics[width=.8\hsize]{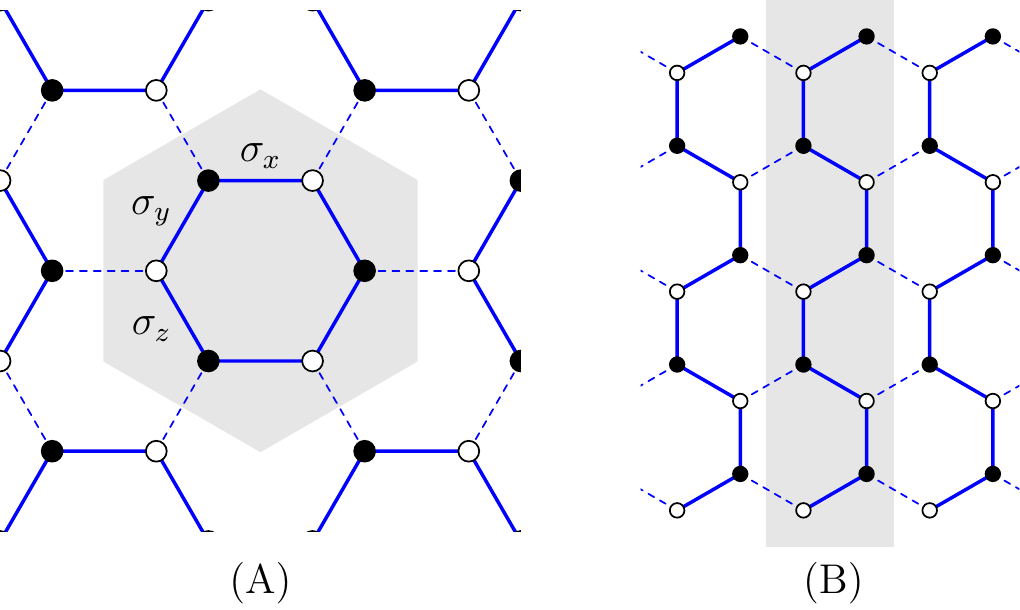}
\caption{(Color online) Clusters used in this work. Left: 6-site cluster used
for the two-dimensional model. Right: 10-site cluster used for one-dimensional
zigzag ribbons. Intercluster links are represented by dashed lines and the
repeated unit is shaded in gray.}
\label{fig:cluster}
\end{figure}

\begin{figure}[htb]
\includegraphics[width=.7\hsize]{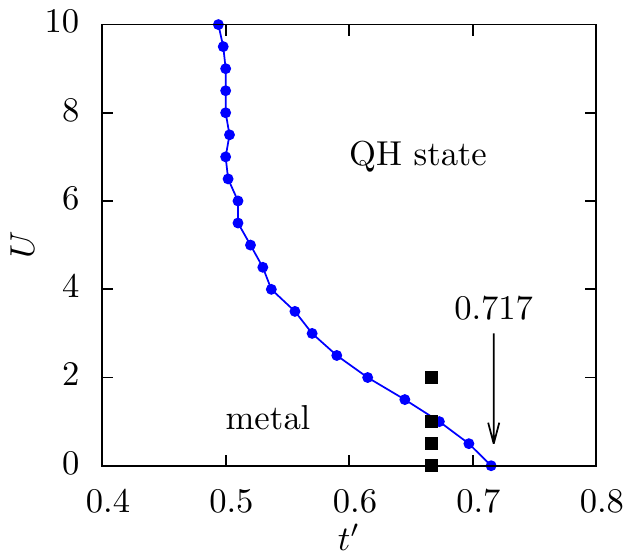} 
\caption{(Color online).
The phase diagram of the model in U-$t'$ plane. The four black squares correspond
to the spectral plots of Fig.~\ref{fig:spectral1}.} 
\label{fig:phasedia}
\end{figure}

\begin{figure}[ht]
\includegraphics[width=\hsize]{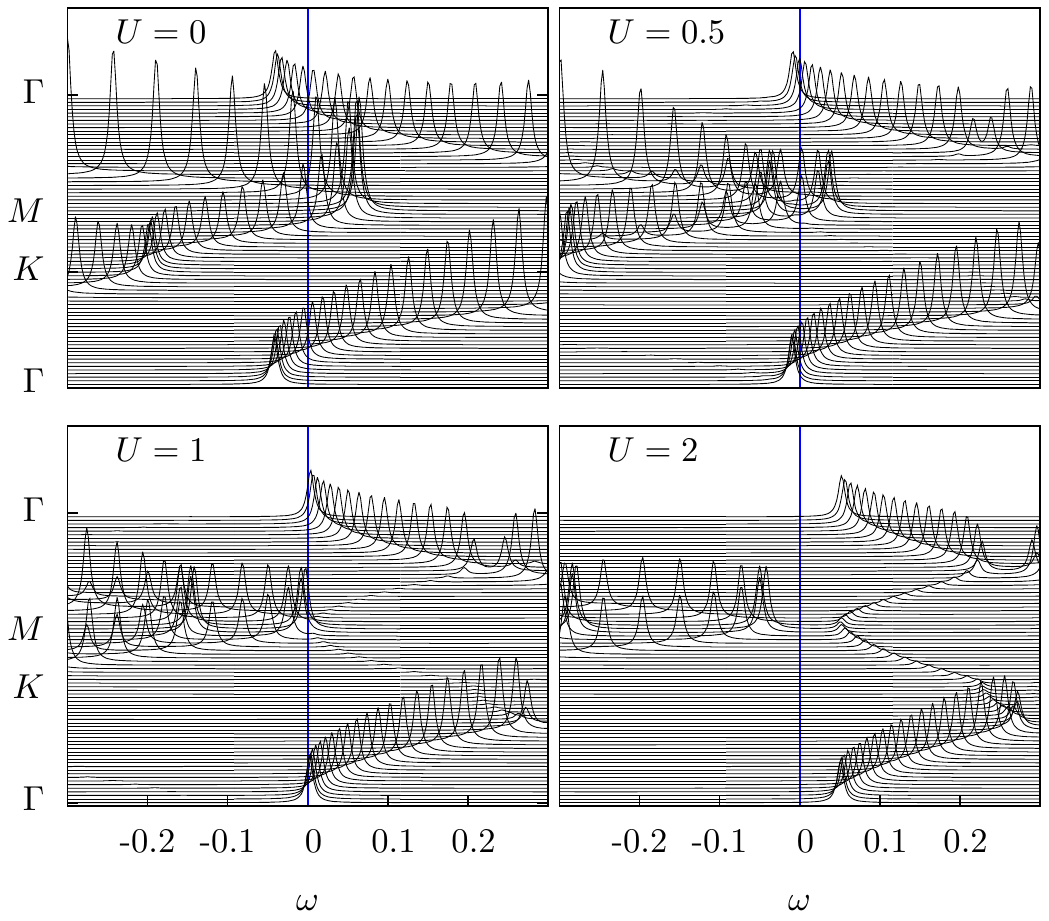}
\caption{Spectral function $A(\kv,\omega)$, as a function of $\omega$, for
spin-up particles and wave vectors along high-symmetry directions at $t'=2/3$, quarter-filling and $U=0,0.5,1$ and 2. The energy unit is set by $t=1$.} 
\label{fig:spectral1}
\end{figure}

\begin{figure}[th]
\includegraphics[width=\hsize]{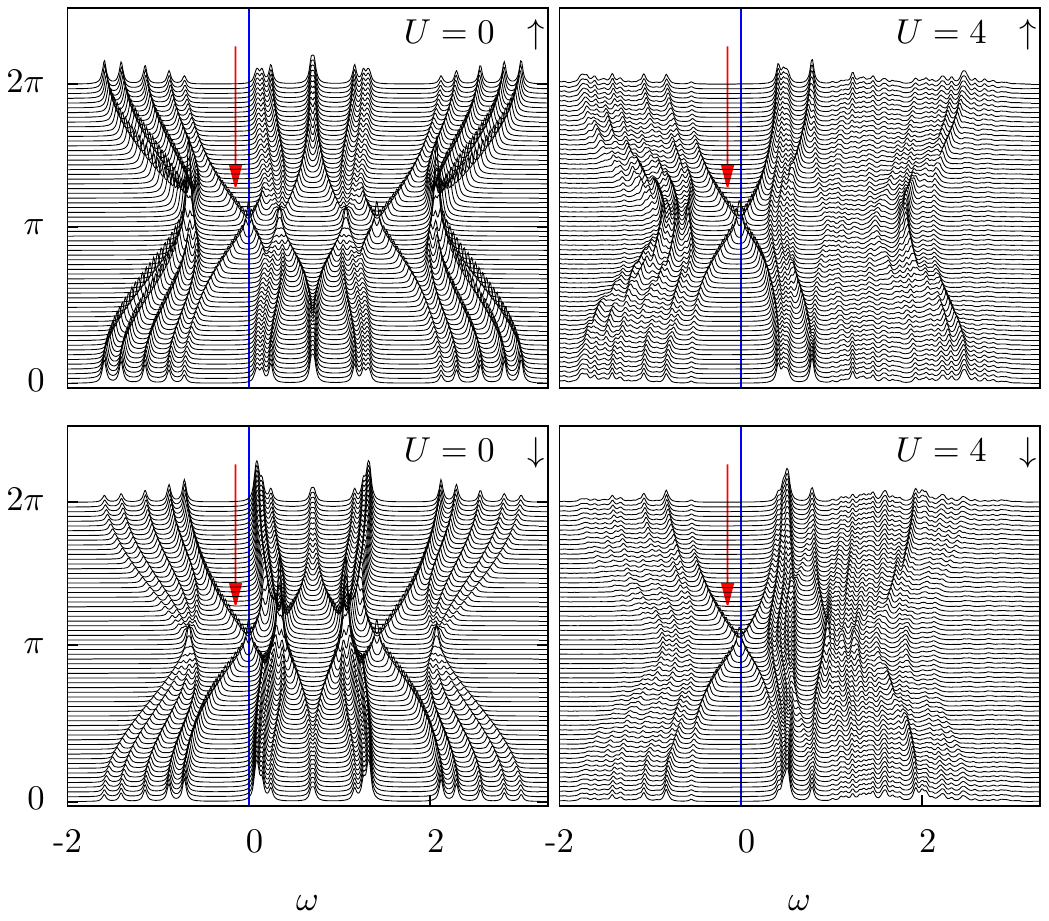}
\caption{(Color online) Top: Single particle spectral function of zigzag ribbons of 10 sites at $t'=t=1$, as a function of energy $\omega$. Top: up spins; bottom: down spins. 
The red arrows indicate that the chiral edge states intersect the Fermi level for both the non-interacting and interacting
systems}
\label{fig:EdgeU}
\end{figure}

We have applied the VCA to the current system by treating the cluster spin
magnetization $M_c$ and the cluster chemical potential $\mu_c$ as variational
parameters.
Fig.~\ref{fig:cluster} illustrates the clusters that were used in this work.
Allowing $\mu_c$ to be different from $\mu$ and adopting the value that 
makes the Potthoff functional $\Omega(M_c,\mu_c)$ stationary ensures 
thermodynamic consistency, i.e., that the electron density $n$ calculated 
from the CPT Green function coincides with 
$-\partial\Omega/\partial\mu$.\cite{Aichhorn:2006rt}
This procedure allows us to calculate a more accurate value of the electron
density $n$ as a function of chemical potential $\mu$, compared with the simple CPT result.
This in turn allows us to better estimate the gap in the one-particle density
of states $N(\omega)$.

We have scanned several values of the interaction $U$ and of the spin-dependent
hopping $t'$, in order to find whether the system remains gapped at
quarter-filling.
Fig.~\ref{fig:phasedia} shows the phase diagram thus obtained, on the $t'-U$
plane.
The curve shown is the phase boundary, i.e., the critical value $t'_c$ at which
the Fermi energy moves from the gap to within the band, as a function of $U$.
In the insulating region (right) the system remains in the quantum Hall state.
Left of the line, the gap disappears and the system becomes metallic.
We call this a {\em chiral metal} because non quantized Hall current flows along
its boundary.
We conclude that the topology of the band is protected for $U>0$, and that a
gapped state exists for $t'\agt 0.5$ if $U$ is large enough.

In the interval from $t'\approx 0.5$ to $t'=0.717$, the quantum Hall state is entered from the chiral metal simply by increasing the interaction strength. This is a Lifshitz transition, as illustrated on Fig.~\ref{fig:spectral1}.
Note that even though a finite Lorenzian broadening has been used to make these plots,
the phase diagram of Fig.~\ref{fig:phasedia} was obtained by using several values of $\eta$ and
extrapolating the density of states towards $\eta=0$.

We also calculated the spectral weight of zizgag ribbons of width 10 at $t'=t$
and quarter filling. The system is shown on Fig.~\ref{fig:cluster}(B) and the
spectral weight on Fig.~\ref{fig:EdgeU}.
The bottom four bands are nearly degenerate at $k=\pi$ in the presence of
interactions and are still clearly separated from the chiral edge state (shown
by the red arrow) near that wave vector.
This indicates that the chiral edge state persists at $U>0$: the interaction
does not destroy the topology of the bands.

\section{Conclusion}

We have shown that time-reversal-breaking spin-dependent
hopping, in a model that can be realized in fermionic cold atom
systems,\cite{Duan:2003dq,
Zhang:2007cr,Dusuel:2008fk,*Vidal:2008uq,Jordens:2008nx} leads to
anomalous quantum Hall states at $\frac14$ and $\frac34$ filling.  These
states are robust against interactions and have a clear experimental
signature: As we showed in section \ref{cnom}, in a confining potential
(as in an optical trap) the PB curvature will make the
atoms rotate. Thus we have a rotating condensate in static trap. This
implies that the atoms have a non-zero orbital angular momentum. So if
the orbital angular momentum can be measured or the rotation detected in
any other way, it would provide an unambiguous signal of
time-reversal symmetry breaking and of the presence of PB curvature 
in the band. We will be describing the effect in more quantitative detail in forthcoming work. 

Note that the absence of time-reversal symmetry and the presence of PB curvature are general features of the model at non-zero $t'$ and are present in both the gapless chiral metal phase and the gapped quantum Hall phase.
The angular momentum also carries a signature of the two phases. 
In the chiral metal phase the orbital angular momentum smoothly increases with the filling factor, as can be seen in Fig..~\ref{fig:ahe}.
By contrast, in the quantum Hall phase, there is a discontinuity at quarter filling. 
This discontinuity is due to the contributions of the edge
states that lie between the bottom two bands. 
The graphs in Fig.~\ref{fig:ahe} have been computed for the infinite lattice in absence of a confining potential. 
However we may expect its effects to persist even in the presence of a trap potential in the form of a sharp increase or a kink in the orbital angular momentum at quarter filling. 
As can be seen from the phase diagram in Fig.~\ref{fig:phasedia}, there is a transition between the chiral metal phase and the quantum Hall phase as $t'$ is increased at
all $U$.  
As we pointed out earlier, the time-reversal symmetry breaking term  $t'$ is induced by the three auxiliary laser beams that create the spin-dependent barriers in the potential. 
Thus $t'$ can be tuned by adjusting the intensity of the three laser beams. 
We have argued that it may be possible to probe the transition that occurs when the auxiliary laser beam intensity is increased by measuring the orbital angular momentum and looking for the appearance of a kink at quarter filling.
Techniques have been developed to measure the angular momentum of bosonic condensates\cite{Chevy:2000fk} and photons in a laser beam.\cite{Lavery:2013uq} 
Our results motivate the search for fractional anomalous quantum Hall states in this model when the bands are partially filled. 

We are grateful to G. Baskaran, H.R. Krishnamurthy and A.-M.S.~Tremblay
for useful discussions.
Computational resources were provided by Compute Canada and Calcul Qu\'ebec.

%

\end{document}